# Using Naïve Bayes Algorithm to Students' bachelor Academic Performances Analysis

[1]Fahad Razaque, [1]Nareena Soomro, [2]Shoaib Ahmed Shaikh,

[1]Department of Computing
Indus University,
[2] Hamdard University,
[3]Aligarh Institute Of Technology
Karachi, Sindh, Pakistan
fahad.indus1337@gmail.com, nainee_soom@yahoo.com,
shoaib.ahmed@hamdard.edu.pk, samojeved@live.com,
natesh_solanki1992@yahoo.com & huma@indus.edu.pk

[4]Safeeullah Soomro, [1]Javed Ahmed Samo, [3]Natesh Kumar and [1]Huma Dharejo

[4]College of Computer Studies

AMA International University

Kingdom of Bahrain

s.soomro@amaiu.edu.bh,

*Abstract*- **Academic Data Mining was one of emerging field which comprise procedure of examined students' details by different elements such as earlier semester marks, attendance, assignment, discussion, lab work were of used to improved bachelor academic performance of students, and overcome difficulties of low ranks of bachelor students. It was extracted useful knowledge from bachelor academic students data collected from department of Computing. Subsequently preprocessing data, which was applied data mining techniques to discover classification and clustering. In this study, classification method was described which was based on naïve byes algorithm and used for Academic data mining. It was supportive to students along with to lecturers for evaluation of academic performance. It was cautionary method for student's to progress their performance of study.**

## I. INTRODUCTION

Many organization of higher education was set up across Pakistan. Conversely education's quality was judge by rate of success student's and to what degree organization was able of preserved students. Predicting performance of student was aid recognized students who be by possibility of failure and therefore management was provided timely assist and obtain important steps to instructor students to improved performance. The capability to predict performance of student was imperative in education sector. Using data mining method which was data in large quantity and discover concealed information sample that was cooperative in decision making. It was identification of different aspects that affected student of learning activities and performance throughout educational sector. Creation of prediction by classification data mining method on foundations of identified predictive keyword.

Humanizing academic performance of students was not simple duty for academic area of higher education. The performance of academic in computing students throughout first year at university was revolving angle in bachelor academic lane and typically intrudes on General Point Average (GPA) in important method. The assessment of students' elements such as assignment, quizzes, midterm and final exam, lab work was studied. It was suggested that correlated information ought to be transmitted to lecturer earlier than transmission of exam. It was helped lecturers to decrease failure percentage to significant level and progress student's performance. In this research based on naïve baye's of data mining technique and data clustering that empowered academia to predicted SGPA (GPA) of students and based on that lecturer was revenue essential stage to improved bachelor academic performance of student. SGPA (Semester Graded Point Average) was usually used pointer of bachelor academic performance which was various departments set least GPA that was been maintained. Consequently, GPA static remained most shared factor used by educational organizer to evaluated evolution in bachelor academic sector. Various factors were performance such as hurdles to student attained and conserved high GPA that mirrored overall performance of academic, throughout tenancy in university. These aspects was been targeted by lecturers in evolving policies to progressed education of student and also performance of academic through method of observing performance evolution. With the help of clustering procedure and naïve Bayes of data mining method it was potential to discovered main characteristics for upcoming prediction. Data clustering was procedure of extracted earlier unidentified, valid, positional beneficial and concealed pattern from bulky data set. The clustering was to divider students into homogeneous clusters according to appearances and skills which were helped both lecturer and student to enrich quality of education. Ensuing every preserved lecture and quiz, lecturers used residual class time as discussion period. It impartial in discussion period was to grow students to talk around material from lecture, discuss answers to quiz, and checked sympathetic. Uncertainty students was not instantly start with questions, so guide lecture slides and focused discussion throughout by lecturers rapid discussion. Every lecturer opinion dissimilar portions of section as more significant based on experience.

It shortage of method education was obvious in tactic industry frequently complained that fresh computer science graduates program was unready for tackling real world software engineering projects. [4, 3, 8, 14].University was also recognized shortage and was attempted to remedy with wide range of innovations designed to create class projects more closely resemble individuals industry. It was included such as deliberately introduced practical difficulties into project, maintained large-scale, ongoing project that different sets of students work on since semester to semester [9], needful students to work on practical project funded by industrial organization [6], integrating multiple universities and disciplines into project [2], and many others.

## II. METHOD AND MATERIAL

The approach using Naïve byes classifier cognitive was suggested for evaluation of student's performance. The elements measured for general evaluation of students was educational, attendance and extra curriculum activities. The information was concealed among academic data set that was extractable through data mining methods. The classification task was used to estimate performance of student and as there are many approaches that are used for data classification, naïve byes classifier was used as included information was extracted that defined performance of student in end semester examination. It was helped in recognized drop-out and students who needed extra care, allowed lecturer to provide appropriate counsel. In academic system, performance of student was resolute by inner assessment and last semester examination. It was carried out using lecturer based upon performance of student in academic actions such as quiz's, discussion, assignments, attendance and lab work. Final semester examination was result by student which was obtained minimum marks to pass semester in final semester examination. The system started from problem definition, then describes data set and preprocessing method executed, and experiments of results, knowledge representation process.

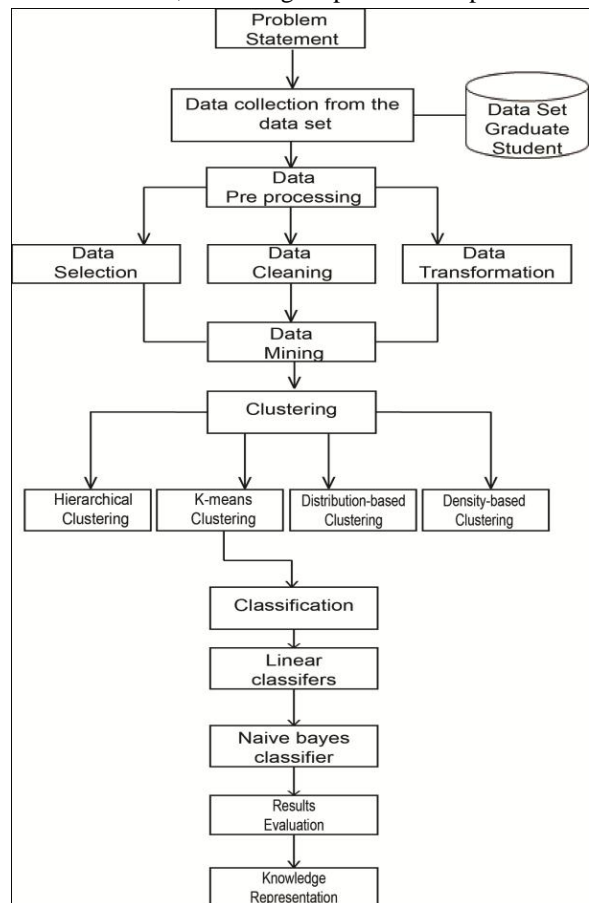

Fig.1 System Frame Work

### A. Data preparation and Data Pre-processing

Data set used was taken student's data from department of computing. The data preparation determination was examined and transformed raw data in order to create them mean more and improved data quality. Without data preparation, hidden information was not easily accessed using data mining models (Pyle, 1999).Data Pre-processing step was executed to develop excellence of data set through removed incomplete values. Data set considered, 61 records were removed from 660 entire data and simply 591 records were prepared for data mining method later. When pre-processing method applied on data set, 91 records were eliminated from data set of 591 records which left only 500 clean records. The total, data of student comprised 160 missing values in numerous parameters from 660 records was ignored from data set. The total numbers of records was reduced to 500.

### B. Data Selection and Transformation

The fields were nominated that was required for data mining which was selected variable, although some of data for variables was mined from data set. The student's performances were deliberated that was used to predict current students' performance in their coming semester.

*C. Data clustering:*

It was arithmetical and unsupervised data investigation method which was classified duplicate information into homogenous cluster. It was used to operate large data set to discover association and concealed pattern helped to make decision efficacy and with rapidly. The cluster analysis was student to segment large data set into subsets referred for instance clusters. Every cluster was group of information items that was related to each other was placed inside corresponding cluster but was unrelated to items in other clusters [13].

*D. Classification:*

It was usually applied method in academic data mining that predicts group exists in data set. Classification was usually applied method in academic data mining that predicts group exists in data set [16]. It was used by scholars in academic arena to better comprehend behaviors of student, to improved instruction ability, and to deliver alternate solution for problems arises in Department of CS and SE[11, 10, 1].Classification was discovered model for predicting academic performance of student to identify students at risk. The semester result in order to predicted CS and SE student's final results.

*E. Naïve Bayes Algorithm:*

The student performance was predicted expending data mining method named classification rules. The NB (Naïve Bayes) classification algorithm was used by administrator to predict student performance in future semester based on earlier semester result and behavior. A Naïve Bayes classifier was simple probabilistic classifier founded on relating Bayes theorem by naive impartiality assumptions. NBC (Naïve Bayes classifiers) was trained extremely expeditiously in supervised education location. It was easy to understand, required training data to parameters estimate, Unresponsive to unrelated features, handled real and distinct data well [15, 5].

## III. EXPERIMENTAL RESULTS

The enormous data stored in academic dataset that contained valuable information for predict performance of students. The classification was used to predict end grade of students and as there are many approaches as data classification, naïve byes in linear classifier method was used here. Using studied and observed each cluster; it was form table determining characteristics of each Cluster and comparison between all clusters as displayed in Table 1 which was naïve byes classifier predicted in percentage of cluster as C1 96.8%, C2 93.9% and C3 98.8% that was best cluster.

TABLE I
PERCENTAGE OF CLUSTERS (C1, C2, C3) USING NAÏVE BYES CLASSIFIER

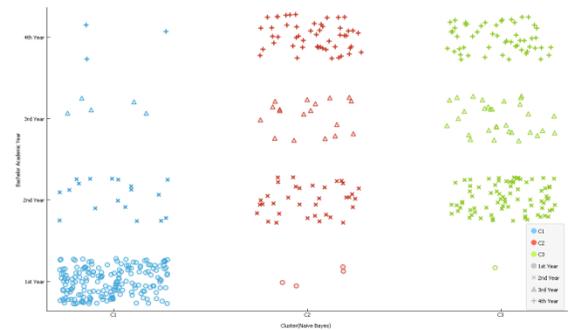

| Actual | C1 | C2 | C3 | Σ |
|---|---|---|---|---|
| C1 | 96.8 % | 0.9 % | 0.6 % | 212 |
| C2 | 1.4 % | 93.9 % | 0.6 % | 112 |
| C3 | 1.8 % | 5.2 % | 98.8 % | 176 |
| Σ | 217 | 115 | 168 | 500 |

Fig 2 depicts graphical representation of clustering as naïve byes classifier was data mining task which discovered strength of student on semesters based that belong to one cluster was more related to each other than to student be applicable to diverse cluster. Impartial of clustering was to find high-quality clusters such that cluster (C1, C2, and C3) in $2^{nd}$ year distances was maximized and (C1, C2, and C3) in $3^{rd}$ year distances was minimized.

Fig.2 Naïve Byes Classifier (Clusters)

The clustering method applied in research was k-means; that experiment was to select best cluster center to be centroid [7]. The clustering method produced with three clusters. Figure 3 indicate resulted centroid Table where from figure that shows average value of each cluster; example cluster labeled Cluster (C1, C2, and C3) have number of students in $1^{st}$ semester of each year was maximum range of GPA into first year(2.82-3.195), second year (2.97-3.029), third year( 2.97-2.98) and fourth (2.96-2.985).

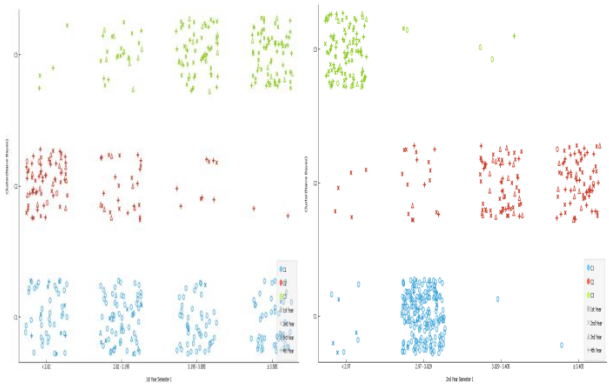

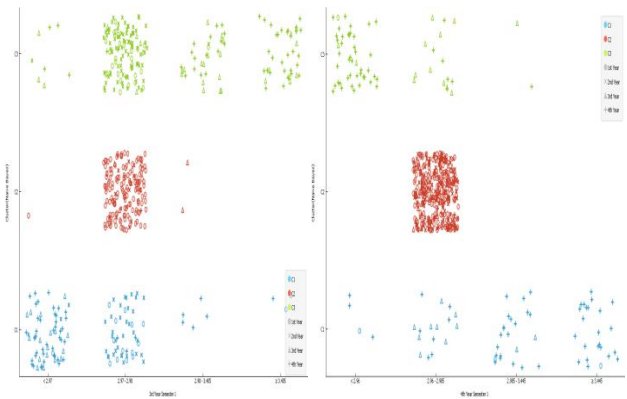

Fig.3 First, second, third, fourth years of first semester student quantity classify through GPA.

Each lecturer used prerecorded lectures for each learning section which was lectures distributed same way to students in each class, with students choose format lecturers wish to use for prerecorded lectures. In order to make sure that students actually read lectures before coming to class, lecturers used weekly quizzes on discussion days that account for 15% of course grade. Fig 4 an illustrated cluster such that cluster (C2) distances was maximized and cluster (C2 and C3) distances was minimized.

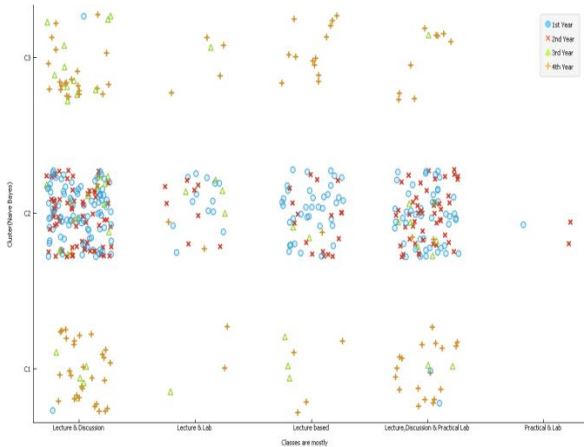

Fig. 4 Academic years classify through (Lecture & discussion).

Prediction that included bachelor academic years was necessitated for effective prediction of student's performance. Prediction's student of performance with high Probability was helpful for identify students with high academic achievements firstly. It was required that identified students were assisted more by lecturer so that performance was improved in future. Figure5shows that Impartial of probability was to find high performance such that bachelor academic year in 1$^{st}$ year probability was high and 3$^{rd}$ year probability was low.

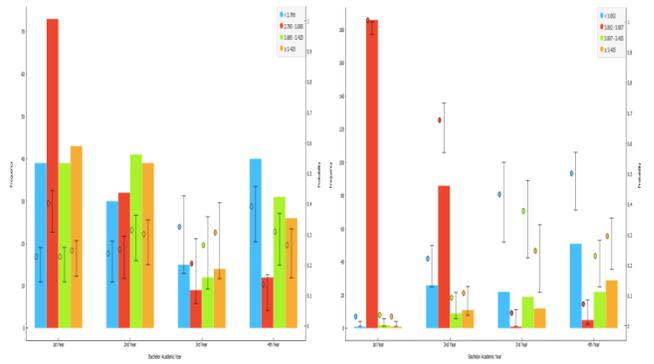

Fig. 5 Predict the performance of student in bachelor academic years using probability

Fig 6 an illustrated student's academic year's probability regarding the performance such as lecture and discussion was high and practical and lab was low probability.

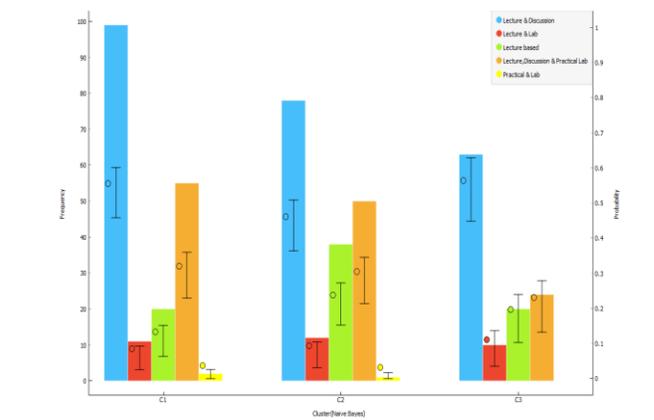

Fig. 6 Academic years classify through (Lecture & discussion).

Fig 7 shows that bachelor academic year were knowledge analysis; mostly classes' attendance was punctuality analysis. Similarly in knowledge and punctuality were combined to performance analysis which was further combined with extra coaching activities to overall ranking of student.

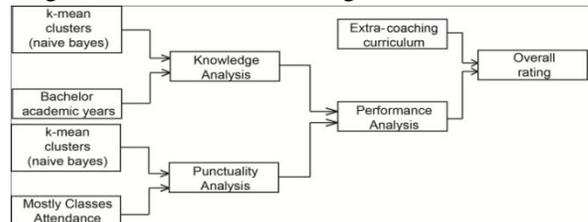

Fig.7 evaluating the overall result on the bases of hierarchy

Fig 8 an illustrated all over ranking such that student have coaching rating node distances was maximized in yes parameter and from university level lecture & discussion node distances was maximized.

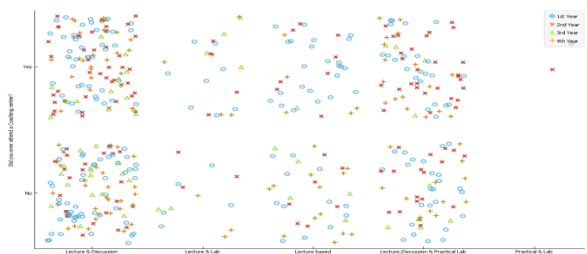

Fig.8 Overall result from evaluation base on coaching, mostly classes in university and bachelor academic year.

Fig 9 showed all over evaluation such that year wise student have coaching ranking and from university level lecture & discussion probability were high in yes parameter.

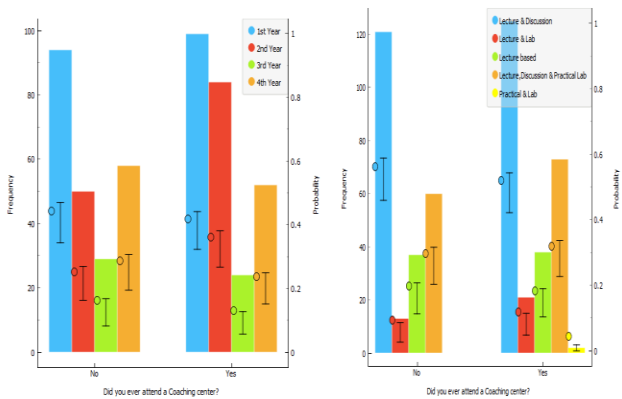

Fig.9 Probability of all over result on the bases of evaluates coaching, bachelor academic and mostly classes.

## IV. CONCLUSION

It was used classification approached which was Naïve Bayesian classifier to predict GPA of graduate student. Also it clustered students into collections using K-Means clustering algorithm. Data such as Attendance, discussion and Assignment grades was collected from student earlier data set, to predict performance at semester end. This research was helped to students and lecturers to improved students result who was at danger of failure. This research was also effort to identify students who required special consideration to decrease failure and taking suitable action for upcoming semester examination.